# Atomic Parameters for Monte Carlo Transport Simulation: Survey, Validation and Induced Systematic Effects

H. Seo, M. G. Pia, L. Quintieri, M. Begalli, P. Saracco and C. H. Kim

*Abstract*–A wide survey has been performed, concerning atomic binding energies and ionization energies used by well-known general purpose Monte Carlo codes and a few specialized electromagnetic models for track structure simulation. Validation results are reported.

## I. INTRODUCTION

The simulation of particle interactions in matters involves a number of atomic physics parameters, like electron binding energies, ionization energies etc. These parameters affect physics quantities, like cross sections of electromagnetic processes and secondary particle spectra; their values can be source of systematic effects in the simulation results.

A wide survey has been performed over well-known Monte Carlo codes and a few specialized electromagnetic models for track structure simulation, concerning atomic binding energies and ionization energies. The set of examined codes includes EGS [1], EGSnrc [2], Geant4 [3][4], MCNP/MCNPX [5][6] and Penelope[7].

Despite the fundamental character of these atomic parameters, there is no consensus among the various Monte Carlo systems and physics models about their values. A variety of experimental and theoretical tabulations of these parameters are used in Monte Carlo codes; in some cases even different values are used in the same Monte Carlo system. The values of these atomic parameters can be source of systematic effects in the simulation results.

The identification of optimal values of these parameters to be used in Monte Carlo simulation is far from trivial. Experimental data often exhibit discrepancies and may be affected by systematic effects. A large scale effort has been invested in the evaluation of the parameters currently used by the various Monte Carlo codes and physics models.

The validation process adopted two complementary approaches: on one side the validation of the parameter values based on direct experimental measurements, on the other side the validation through experimental comparisons of related physics observables, depending on these parameters. Epistemic uncertainties are present in Monte Carlo codes, when the quality of experimental data prevents the achievement of firm conclusions regarding the correct values of such parameters.

A related study concerning the accuracy of radiative transition probability calculations is described in [8]; the results are not reported here.

The outcome of these studies is relevant to optimize the accuracy of Monte Carlo codes and of the data libraries they use.

## II. COMPILATIONS OF ATOMIC BINDING ENERGIES

A set of tabulations of atomic binding energies has been evaluated; they are exploited by widely used Monte Carlo systems and in experimental practice. They include:
- the compilation by Carlson [9], used by MCNPX and Penelope 2008 version
- the compilation by Lotz [10], on which Carlson's compilation was largely based
- the compilations in the Table of Isotopes [11][12], 1978 and 1996 editions, respectively used by EGSnrc (and formerly by EGS4) and EGS5; the binding energies of the earlier edition are also used in the simulation of Doppler broadening in Geant4 low energy electromagnetic package
- the tabulation in the Evaluated Data Library (EADL) [13], used by Geant4 low energy electromagnetic package
- the set of binding energy values hard-coded in the G4AtomicShells class of Geant4 materials package, which are nominally based on Carlson's compilation
- the compilation in the X-ray Data Booklet [14].

The compilations by Carlson and Lotz, and EADL tabulations are the result of theoretical calculations; the compilations in the Table of Isotopes and in the X-ray Data Booklet are of experimental origin.

## III. VALIDATION RESULTS

The accuracy of binding energy calculations has been estimated with respect to reproducing the experimental X-ray energies reported in the review by DesLattes et al. [15] and other high precision experimental measurements of atomic binding energies. Only the first part of the study is reported in this paper.

M. G. Pia and P. Saracco are with INFN Sezione di Genova, Via Dodecaneso 33, I-16146 Genova, Italy (telephone: +39 010 3536328, e-mail: MariaGrazia.Pia@ge.infn.it).

L. Quintieri is with INFN Laboratori Nazionali di Frascati, Frascati, Italy (e-mail: Lina.Quintieri@lnf.infn.it).

H. Seo and C. H. Kim are with the Department of Nuclear Engineering, Hanyang University, Seoul 133-791, Korea (e-mail: shee@hanyang.ac.kr; chkim@hanyang.ac.kr).

M. Begalli is with State University of Rio de Janeiro, Brazil (e-mail: Marcia.Begalli@cern.ch).

It is worthwhile to remind the reader that characteristic X-ray energies are determined by the difference of the binding energies associated with the subshells involved in a radiative transition.

A comparison of X–ray energies based on EADL with respect to the same data was previously performed and is documented in [16]; this study highlighted that other compilations achieve better accuracy at reproducing experimental X-ray energies.

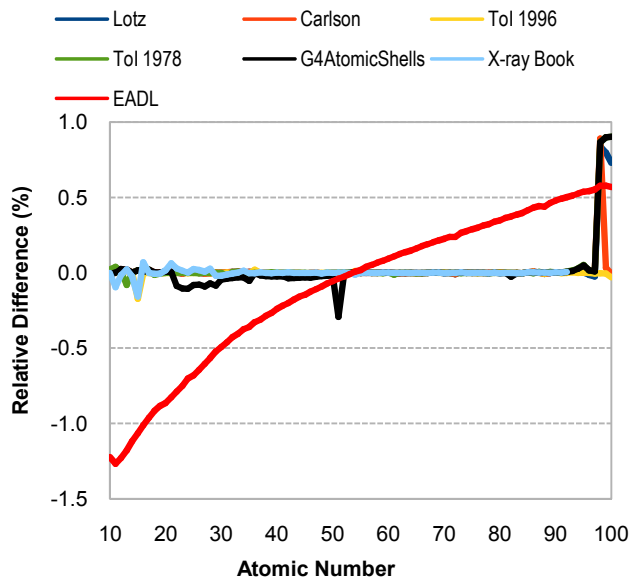

Fig. 1. $KL_3$ radiative transition: relative difference of X-ray energies resulting from various binding energies compilations with respect to experimental data in [15].

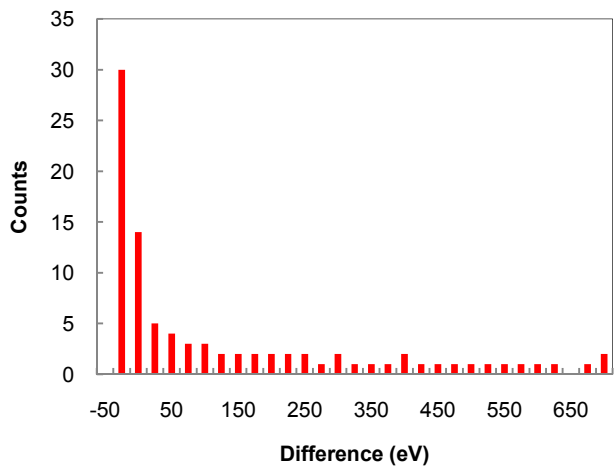

Fig. 2. $KL_3$ radiative transition: absolute difference of X-ray energies derived from EADL binding energies with respect to experimental data in [15].

The analysis was based on rigorous statistical methods; only qualitative plots highlighting some preliminary significant results are reported here, while the full set of results, including the detailed analysis that produced them, is meant to be documented in a following dedicated paper.

Fig. 1 to Fig. 6 show the relative and absolute differences of X-ray energies derived from various binding energy compilations with respect to experimental data, respectively for $KL_3$ and $L_1M_3$ transitions as an example. One can observe that, while relative differences with respect to experiment appear to be of the order of 1-2% at most for all compilations, some binding energy tabulations produce very accurate estimates of X-ray energies, differing from measured values by a few electronvolts only, X-ray energies deriving from EADL binding energies may differ even some hundreds of electronvolts with respect to experimental references.

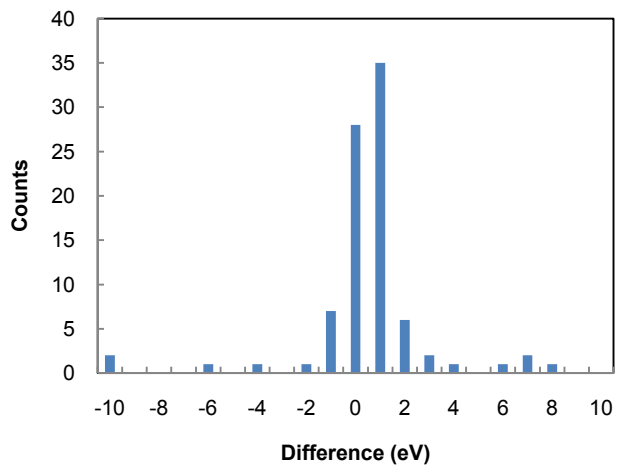

Fig. 3. $KL_3$ radiative transition: absolute difference of X-ray energies derived from Carlson's binding energies with respect to experimental data in [15].

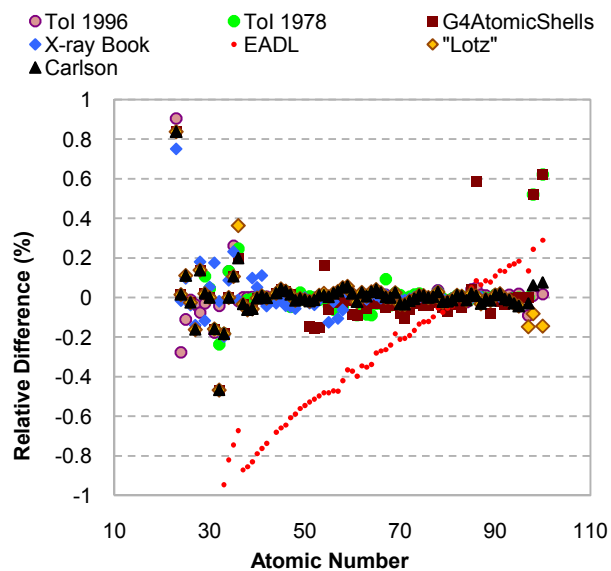

Fig. 4. $L_1M_3$ radiative transition: relative difference of X-ray energies resulting from various binding energies compilations with respect to experimental data in [15].

Ionization energies reported in various compilations have been compared to the reference experimental data in NIST's Physical Reference Data collection. The relative differences corresponding to three compilations are shown in Fig. 7.

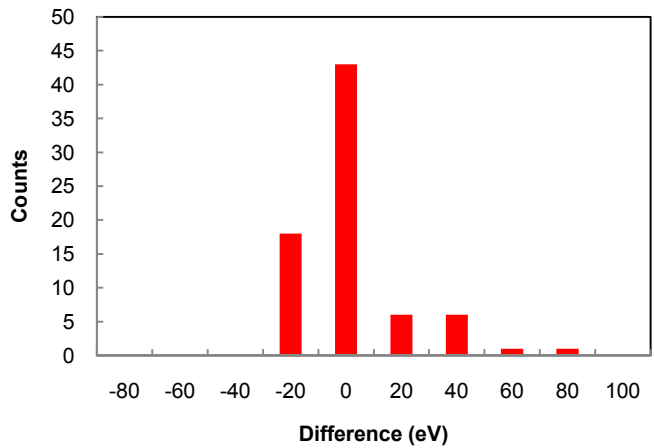

Fig. 5. $L_1M_3$ radiative transition: absolute difference of X-ray energies derived from EADL binding energies with respect to experimental data in [15].

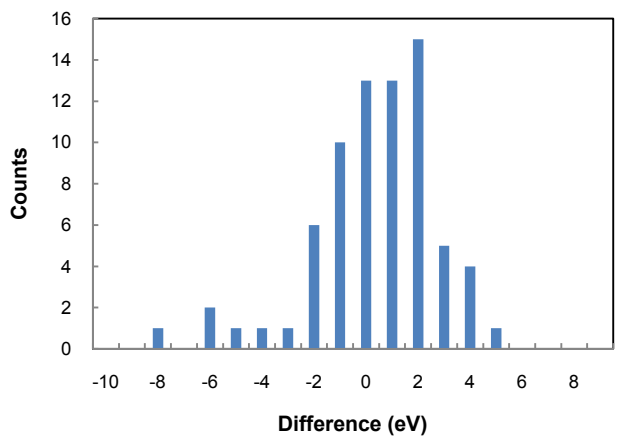

Fig. 6. $L_1M_3$ radiative transition: absolute difference of X-ray energies derived from Carlson binding energies with respect to experimental data in [15].

## IV. EFFECTS ON OTHER PHYSICS QUANTITIES

Some calculations of physics quantities typically used in Monte Carlo codes, like cross sections, involve atomic parameters; inaccurate values of these parameters may be responsible for systematic effects in the simulation.

An example of such possible effects is illustrated in Fig. 8. The plot shows the electron ionization cross section as a function of energy calculated by the Binary-Encounter-Bethe (BEB) model, whose formulation involves binding energies and average electron kinetic energies associated with the relevant orbitals. Significant differences are visible, when EADL or Lotz ionization energies are used in the calculation, while different inner shell binding energies or electron kinetic energies appear to have relatively small effects.

The discrimination of which compilations of ionization energies produce more accurate cross sections is not straightforward: as one can see in Fig. 9 and Fig. 10, for example, one can identify cases where either EADL or Lotz ionization energies produce cross sections consistent with experimental data, while in either cases, like in Fig. 11, inconsistencies in cross section measurements themselves do not allow firm conclusions. One can draw conclusions only based on a thorough statistical analysis over a large data sample.

The results of this analysis will be documented in a paper currently in preparation.

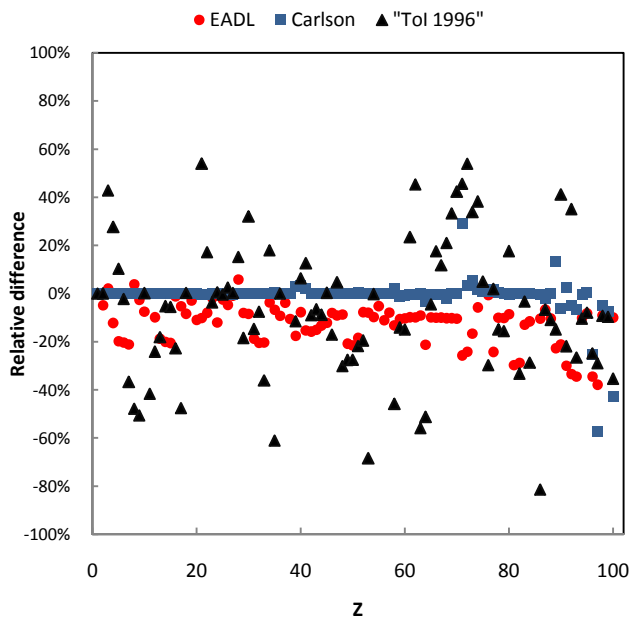

Fig. 7. Ionization energies: relative difference of values tabulated in three compilations (EADL, Carlson and Table of Isotopes 1996 edition) with respect to NIST's reference experimental data.

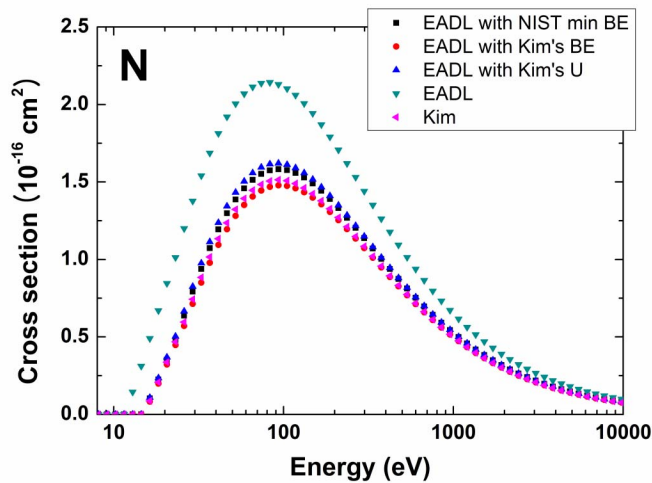

Fig. 8. BEB cross section for nitrogen ionization as a function of electron energy: values based on EADL inner shell binding energies and electron kinetic energies and NIST ionization energy (black squares), on EADL inner shell binding energies and electron kinetic energies, ionization energy as in [18] (red circles), on EADL inner shell binding energies, NIST ionization energy and electron kinetic energies as in [18] (blue triangles), on all EADL parameters (green triangles) and on parameters all as in [18].

## V. CONCLUSION

A wide survey of atomic binding energies compilations in the literature has been performed. Comparisons of experimental physics observables depending on these quantities show that, among the evaluated compilations, EADL exhibits relatively worse accuracy than other tabulations analyzed in this study. The set of binding energies implemented in the Geant4 G4AtomicShells class also appears less accurate than other tabulations.

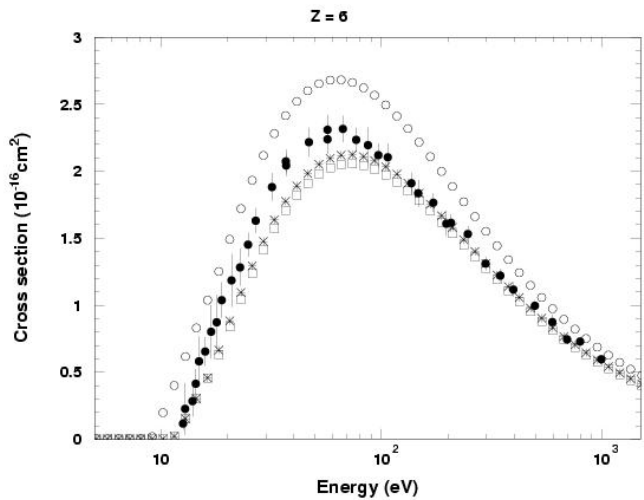

Fig. 9. BEB cross section for carbon ionization as a function of electron energy: values based on EADL inner shell binding energies and ionization energy (empty circles), on EADL inner shell binding energies and NIST ionization energy (empty squares), on Lotz binding energies (crosses) and experimental data [19] (filled symbols).

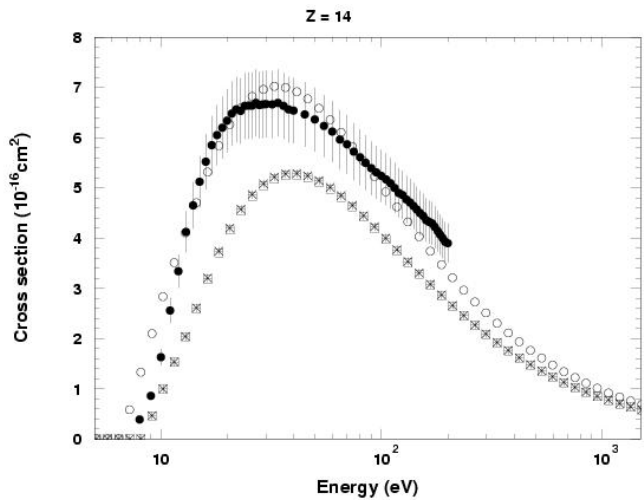

Fig. 10. BEB cross section for silicon ionization as a function of electron energy: values based on EADL inner shell binding energies and ionization energy (empty circles), on EADL inner shell binding energies and NIST ionization energy (empty squares), on Lotz binding energies (crosses) and experimental data [20] (filled symbols).

Other recent tests of radiative transition probabilities have shown the need of updating EADL to achieve better accuracy. Sources for such an improvement have been identified.

It is worthwhile to remind Monte Carlo simulation developers that caution should be exercised in Monte Carlo codes using EADL data, if one intends to upgrade atomic parameters derived from EADL to more accurate values; by changing the atomic parameters derived from the current version of EADL, although not representing the state-of-the-art, one risks possible inconsistencies in other parts of the code using associated EEDL [23] and EPDL [24] data libraries. A coordinated effort aimed at the validation and, if necessary, improvement of these three related data libraries would be desirable.

The complete set of results of this study will be documented and discussed in depth in dedicated papers.

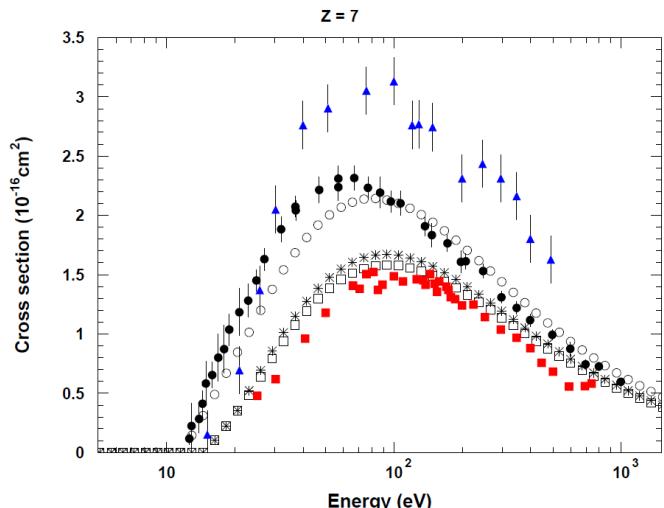

Fig. 11. BEB cross section for nitrogen ionization as a function of electron energy: values based on EADL inner shell binding energies and ionization energy (empty circles), on EADL inner shell binding energies and NIST ionization energy (empty squares), on Lotz binding energies (crosses) and experimental data [19][21][22] (filled symbols).

ACKNOWLEDGMENT

The authors express their gratitude to CERN for support to the research described in this paper.

The authors thank Tullio Basaglia and the CERN Library for essential support to this study.